\documentclass[conference]{IEEEtran}

\usepackage{cite}
\usepackage{graphicx}
\usepackage{xcolor}
\usepackage{pgfplots}
\usepackage{multirow}
\usepackage{upgreek}
\usepackage{amssymb}
\usepackage{amsmath, mathtools}
\usepackage{cases}
\usepackage{pifont}
\usepackage{bm}
\usepackage{amsthm}

\usepackage{tabularx}
\usepackage{booktabs}
\newcolumntype{Y}{>{\centering\arraybackslash}X}
\usepackage{array}
\usepackage{float}
\usepackage{xcolor}
\usepackage{algorithmic}
\usepackage[linesnumbered,ruled,vlined]{algorithm2e}
\usepackage{changes}
\usepackage{bbm}
\usepackage{soul}
\usepackage{physics}
\usetikzlibrary{arrows,positioning,shapes.geometric,spy}
\usepackage{caption,subcaption}

\pgfplotsset{compat=1.17}

\newcommand{\fixme}[2]{\ifx&#2&{\leavevmode\color{red}#1}\else{\leavevmode\color{red}FIXME\{}#1{\leavevmode\color{red}\}}\footnote{{\leavevmode\color{red}#2}}\PackageWarning{Fixme}{#1: #2}\fi}

\SetCommentSty{mycommfont}

\DeclareMathOperator*{\sgn}{sgn}

\begin{document}

\title{A Tree Search Approach for Maximum-Likelihood Decoding of Reed-Muller Codes}

\author{
\IEEEauthorblockN{Seyyed Ali Hashemi\IEEEauthorrefmark{1}, Nghia Doan\IEEEauthorrefmark{2}, Warren J. Gross\IEEEauthorrefmark{2}, John Cioffi\IEEEauthorrefmark{1}, Andrea Goldsmith\IEEEauthorrefmark{3}}

\IEEEauthorblockA{\IEEEauthorrefmark{1}Department of Electrical Engineering, Stanford University, USA} 
\IEEEauthorblockA{\IEEEauthorrefmark{2}Department of Electrical and Computer Engineering, McGill University, Canada}
\IEEEauthorblockA{\IEEEauthorrefmark{3}Department of Electrical and Computer Engineering, Princeton University, USA}
\IEEEauthorblockA{ahashemi@stanford.edu,
nghia.doan@mail.mcgill.ca,
warren.gross@mcgill.ca,
cioffi@stanford.edu,
goldsmith@princeton.edu}
}

\maketitle

\begin{abstract}
A low-complexity tree search approach is presented that achieves the maximum-likelihood (ML) decoding performance of Reed-Muller (RM) codes. The proposed approach generates a bit-flipping tree that is traversed to find the ML decoding result by performing successive-cancellation decoding after each node visit. A depth-first search (DFS) and a breadth-first search (BFS) scheme are developed and a log-likelihood-ratio-based bit-flipping metric is utilized to avoid redundant node visits in the tree. Several enhancements to the proposed algorithm are presented to further reduce the number of node visits. Simulation results confirm that the BFS scheme provides a lower average number of node visits than the existing tree search approach to decode RM codes.
\end{abstract}

\IEEEpeerreviewmaketitle

\section{Introduction}

The Reed-Muller (RM) code family \cite{Muller,Reed} is one of the oldest channel coding schemes. It has regained attention recently due to its similarity with polar codes \cite{arikan}. Both RM and polar codes are constructed by selecting rows of a generator matrix. The row selection for RM codes is such that the minimum distance among all the codewords is maximized, while the row selection for polar codes minimizes the error probability when the low-complexity successive cancellation (SC) decoder is used. Unlike polar codes, RM codes have the advantage of channel-independent code construction that is particularly beneficial when they are used in rapidly changing environments, e.g. in mmWave and THz bands, which are being considered for 5G and other wireless systems. In addition, RM codes can achieve the capacity of binary erasure channels under maximum-likelihood (ML) decoding \cite{kudekar}. Since ML decoding is generally impractical, RM codes are often decoded using decoders that can approach ML decoding performance, such as recursive list decoding \cite{Dumer06}, decoding on several factor graph permutations \cite{key2010reed,Ali_SP,Kamenev19}, decoding with minimum-weight parity-checks \cite{santi}, and recursive projection-aggregation decoding \cite{Ye20,fathollahi2020sparse}.

SC decoding was first used to decode RM codes in \cite{dumer_SC}. However, the error-correction performance of RM codes under SC decoding is far from satisfactory. It was observed in \cite{yuan_ML} that the resulting error probability of each information bit under SC decoding can be used to perform a bit-flipping tree search to develop a SC ordered search (SCOS) decoding algorithm that achieves ML decoding performance with adaptive complexity. The advantage of the SCOS decoding algorithm over the bit-flipping decoding algorithms for polar codes \cite{SCF,DSCF} is that it does not require a cyclic redundancy check (CRC) to determine if the decoding has succeeded. However, the tree search algorithm in \cite{yuan_ML} relies on two costly computations: the calculation of a metric that has to be performed using empirical methods; and the sorting of the metrics to determine which branch of the tree is explored next.

In this paper, a modified tree search decoding algorithm is presented that, unlike the SCOS decoder in \cite{yuan_ML}, does not require the empirical calculation of a metric, and does not need the sorting operation to determine the next branch traversal. A bit-flipping tree is developed and two tree search algorithms are used to find the ML decoding result: a depth-first search (DFS) approach that has a simple recursive structure; and a breadth-first search (BFS) approach that obtains a lower average decoding complexity than the DFS method. The key idea in the proposed method is to follow the natural bit index order rather than sorting the bit error probabilities. Simulation results show that the proposed methods achieve ML decoding performance while the BFS method requires fewer tree node visits than the SCOS decoder. To further reduce the average and worst-case decoding complexity of the DFS and the BFS schemes, several enhancements are proposed: 1) the depth of the tree is limited to a certain predefined value that provides near-ML decoding performance while significantly reducing the average and worst-case decoding complexity; 2) redundant node visits are eliminated to reduce the worst-case decoding complexity; and 3) a metric is utilized to order the node visits that reduces the average decoding complexity.

\section{Preliminaries}

\subsection{Reed-Muller Codes}

A RM code of length $N=2^m$ and order $r$, ($0 \leq r \leq m$) is denoted as $\mathcal{RM}(r,m)$.
The dimension (number of information bits) of $\mathcal{RM}(r,m)$ is $K=\sum_{i=0}^{r} \binom{m}{i}$ and it has a minimum distance ${d=2^{m-r}}$.
RM codes can be constructed by converting the input bits $\bm{u} = \{u_0,u_1,\ldots,u_{N-1}\}$ to the encoded bits $\bm{x} = \{x_0,x_1,\ldots,x_{N-1}\}$ by a linear transformation as $\bm{x} = \bm{u}\bm{G}^{\otimes m}$, where $\bm{G}^{\otimes m}$ is the $m$-th Kronecker power of the matrix $\bm{G}=\bigl[\begin{smallmatrix} 1&0\\ 1&1 \end{smallmatrix} \bigr]$ \cite{Arikan10}. The construction of RM codes involves generating two sets based on $w_i$, the weight of the $i$-th row of $\bm{G}^{\otimes m}$: a frozen set $\mathcal{F} = \{i|0\leq i < N, w_i < d\}$, where $u_i = 0$ $\forall i \in \mathcal{F}$; and an information set $\mathcal{I} = \{i|0\leq i < N, w_i \ge d\}$, where $u_i$ $\forall i \in \mathcal{I}$ carries the information bits. Both of the sets $\mathcal{I}$ and $\mathcal{F}$ are known to the transmitter and the receiver.

This paper considers binary phase-shift keying (BPSK) modulation on the vector $\bm{x}$ and an additive white Gaussian noise (AWGN) channel model. Thus the received signal is ${\bm{y}=(\mathbf{1}-2\bm{x})+\bm{z}}$, where $\mathbf{1}$ is an all-one vector of size $N$, and $\bm{z} \in \mathbbm{R}^N$ is the Gaussian noise vector with zero mean and variance $\sigma^2$. The log-likelihood ratio (LLR) values associated with the received vector $\bm{y}$
can be calculated as ${\bm{\alpha}_m=\ln{\frac{\mathbb{P}(\bm{x}=0|\bm{y})}{\mathbb{P}(\bm{x}=1|\bm{y})}}=\frac{2\bm{y}}{\sigma^2}}$.

\subsection{Successive-Cancellation Decoding and Its Variants}

SC decoding follows a bit-by-bit decoding schedule. Each bit $u_i$ is estimated as the ML decision based on $\bm{\alpha}_m$, considering all of the previous bits $u_0,\ldots,u_{i-1}$ are decoded correctly and all of the future bits $u_{i+1},\ldots,u_{N-1}$ are unknown. If $i\in\mathcal{F}$, then $\hat{u}_i$, the estimation of $u_i$, is set to $0$. Otherwise, the LLR value of $u_i$ is calculated as
\begin{equation}\label{equ:LLR}
    \alpha_{i} = \ln \frac{\mathbb{P}(u_i=0|\bm{\alpha}_m,\hat{u}_0,\ldots,\hat{u}_{i-1})}{\mathbb{P}(u_i=1|\bm{\alpha}_m,\hat{u}_0,\ldots,\hat{u}_{i-1})}\text{,}
\end{equation}
and $u_i$ is estimated as $0$ if $\alpha_i \geq 0$, and as $1$ otherwise.
To reduce the complexity in the implementation of the SC decoding algorithm, the min-sum approximation \cite{leroux} is used to carry out (\ref{equ:LLR}).

SC decoding is well-suited to work with polar codes of large lengths. However, since SC decoding does not use any available information about the future bits, its error-correction performance on polar codes of moderate lengths and RM codes is not satisfactory. In fact, it was shown in \cite{arikan} that RM codes are asymptotically unreliable under SC decoding. Therefore, variants of SC decoding such as SC list (SCL) decoding and SC flip (SCF) decoding are used for polar codes and RM codes \cite{tal_list,SCF,DSCF,Dumer06}.

SCL decoding is based on running multiple SC decoders in parallel by estimating each information bit as either $0$ or $1$. To avoid exponential growth in the complexity of SCL decoding, at each bit estimation step, only the $L$ most reliable candidate paths are kept in a list. Therefore, SCL decoding with list size $L=2^K$ guarantees ML decoding performance. A path metric is used to select the $L$ most reliable candidates at each step that can be well approximated as \cite{Alexios_LLR_SCLD}
\begin{equation}
	\label{equ:polar:PM_LLR}
	\text{PM}_{i} = \sum_{0}^{i} \beta_i \abs{\alpha_{i}} \text{,}
\end{equation}
where $\alpha_{i}$ is the LLR value of $u_i$ and
\begin{equation*}
    \beta_i = \begin{cases}
		1 & \text{if } \hat{u}_i \neq \frac{1-\sgn(\alpha_i)}{2} \text{,}\\
		0 & \text{otherwise.}
	\end{cases}
\end{equation*}
When $u_{N-1}$ is decoded, the path with the smallest path metric is selected as the decoding result.

SCF decoding is developed for polar codes and it relies on an outer CRC code to detect if the initial SC decoding attempt succeeds or not. In case of a failure, the SCF decoder sequentially flips one bit at a time and performs SC decoding in an attempt to pass the CRC test. In \cite{SCF}, the absolute value of the LLR value for each $u_i$ is selected as a metric to determine the bits that need to be flipped. It was shown in \cite{DSCF} that a dynamic SCF decoding can perform multiple bit-flips at a time. An enhanced bit-flip selection metric based on the LLR value of previous bits is developed that improved the performance of dynamic SCF decoding. However, the dependence of SCF decoding on a CRC renders it unattractive for decoding RM codes.

The SCOS decoding \cite{yuan_ML} achieves the ML decoding performance by following a sequential schedule that starts with a round of SC decoding. The result of SC decoding is used as a potential ML candidate and the path metric associated with this candidate is stored as $\text{PM}_{\text{best}}$ using (\ref{equ:polar:PM_LLR}). Then, the estimated values of the information bits $\hat{u}_i$, $i \in \mathcal{I}$, are flipped as $\hat{u}_i \oplus 1$ and their path metrics are calculated considering all the previous bit estimations are unchanged. Then SC decoding is performed to generate a new candidate. If the path metric of the new candidate is smaller than $\text{PM}_{\text{best}}$, the candidate is stored as the potential ML candidate and $\text{PM}_{\text{best}}$ is updated. To reduce the number of bit-flips, SCOS decoding uses the fact that the path metric calculation in (\ref{equ:polar:PM_LLR}) is non-decreasing in $i$. Therefore, if the path metric associated with a bit-flip is larger than $\text{PM}_\text{best}$, that bit-flip and any subsequent bit-flips cannot result in the ML candidate. SCOS decoding thus creates a list of candidates whose path metrics are less than $\text{PM}_\text{best}$. In order to determine which bit is flipped next in the list of candidates, a score parameter is calculated for each $\hat{u}_i$ as
\begin{equation}
    S_{i} = \text{PM}_{i} + \sum_{j=0}^{i} \ln (1-p_j) \text{,}
    \label{equ:score}
\end{equation}
where $p_j$ is the probability that in SC decoding, the first error occurs in $u_j$. After each SC decoding, the candidates in the list are sorted based on their score and the bit with the minimum score is flipped. To provide a complexity-performance trade-off, SCOS decoding allows for limiting the list size at the cost of no longer guaranteeing ML decoding performance.

\begin{figure*}[t]
    \centering
    \begin{subfigure}{.48\textwidth}
        \begin{tikzpicture}[scale=.94, thick]

\draw (0,0) circle [radius=.3] node {SC};

\draw (0,-.3) -- (-1,-1.2) node [midway,fill=white] {$9$};
\draw (0,-.3) -- (-3,-1.2) node [midway,fill=white] {$1$};
\draw (0,-.3) -- (1,-1.2) node [midway,fill=white] {$13$};
\draw (0,-.3) -- (2.5,-1.2) node [midway,fill=white] {$15$};

\draw (-3.5,-1.2) rectangle ++(1,-.5) node [pos=.5] {$\hat{u}_3 \!\oplus\! 1$};
\draw (-3,-2) circle [radius=.3] node {SC};
\draw (-1.5,-1.2) rectangle ++(1,-.5) node [pos=.5] {$\hat{u}_5 \!\oplus\! 1$};
\draw (-1,-2) circle [radius=.3] node {SC};
\draw (.5,-1.2) rectangle ++(1,-.5) node [pos=.5] {$\hat{u}_6 \!\oplus\! 1$};
\draw (1,-2) circle [radius=.3] node {SC};
\draw (2,-1.2) rectangle ++(1,-.5) node [pos=.5] {$\hat{u}_7 \!\oplus\! 1$};
\draw (2.5,-2) circle [radius=.3] node {SC};

\draw (-3,-2.3) -- (-3.5,-3.2) node [midway,fill=white] {$6$};
\draw (-3,-2.3) -- (-4.6,-3.2) node [midway,fill=white] {$2$};
\draw (-3,-2.3) -- (-2.4,-3.2) node [midway,fill=white] {$8$};

\draw (-1,-2.3) -- (-1,-3.2) node [midway,fill=white] {$10$};
\draw (-1,-2.3) -- (.1,-3.2) node [midway,fill=white] {$12$};

\draw (1,-2.3) -- (1.5,-3.2) node [midway,fill=white] {$14$};

\draw (-5.1,-3.2) rectangle ++(1,-.5) node [pos=.5] {$\hat{u}_5 \!\oplus\! 1$};
\draw (-4.6,-4) circle [radius=.3] node {SC};
\draw (-4,-3.2) rectangle ++(1,-.5) node [pos=.5] {$\hat{u}_6 \!\oplus\! 1$};
\draw (-3.5,-4) circle [radius=.3] node {SC};
\draw (-2.9,-3.2) rectangle ++(1,-.5) node [pos=.5] {$\hat{u}_7 \!\oplus\! 1$};
\draw (-2.4,-4) circle [radius=.3] node {SC};

\draw (-1.5,-3.2) rectangle ++(1,-.5) node [pos=.5] {$\hat{u}_6 \!\oplus\! 1$};
\draw (-1,-4) circle [radius=.3] node {SC};
\draw (-.4,-3.2) rectangle ++(1,-.5) node [pos=.5] {$\hat{u}_7 \!\oplus\! 1$};
\draw (.1,-4) circle [radius=.3] node {SC};

\draw (1,-3.2) rectangle ++(1,-.5) node [pos=.5] {$\hat{u}_7 \!\oplus\! 1$};
\draw (1.5,-4) circle [radius=.3] node {SC};

\draw (-4.6,-4.3) -- (-5,-5.2) node [midway,fill=white] {$3$};
\draw (-4.6,-4.3) -- (-3.9,-5.2) node [midway,fill=white] {$5$};

\draw (-3.5,-4.3) -- (-2.5,-5.2) node [midway,fill=white] {$7$};

\draw (-1,-4.3) -- (-1,-5.2) node [midway,fill=white] {$11$};

\draw (-5.5,-5.2) rectangle ++(1,-.5) node [pos=.5] {$\hat{u}_6 \!\oplus\! 1$};
\draw (-5,-6) circle [radius=.3] node {SC};
\draw (-4.4,-5.2) rectangle ++(1,-.5) node [pos=.5] {$\hat{u}_7 \!\oplus\! 1$};
\draw (-3.9,-6) circle [radius=.3] node {SC};

\draw (-3,-5.2) rectangle ++(1,-.5) node [pos=.5] {$\hat{u}_7 \!\oplus\! 1$};
\draw (-2.5,-6) circle [radius=.3] node {SC};

\draw (-1.5,-5.2) rectangle ++(1,-.5) node [pos=.5] {$\hat{u}_7 \!\oplus\! 1$};
\draw (-1,-6) circle [radius=.3] node {SC};

\draw (-5,-6.3) -- (-5,-7.2) node [midway,fill=white] {$4$};

\draw (-5.5,-7.2) rectangle ++(1,-.5) node [pos=.5] {$\hat{u}_7 \!\oplus\! 1$};
\draw (-5,-8) circle [radius=.3] node {SC};

\end{tikzpicture}
        \caption{DFS.}
        \label{fig:treeDFS}
    \end{subfigure}
    \begin{subfigure}{.48\textwidth}
        \begin{tikzpicture}[scale=.94, thick]

\draw (0,0) circle [radius=.3] node {SC};

\draw (0,-.3) -- (-1,-1.2) node [midway,fill=white] {$2$};
\draw (0,-.3) -- (-3,-1.2) node [midway,fill=white] {$1$};
\draw (0,-.3) -- (1,-1.2) node [midway,fill=white] {$3$};
\draw (0,-.3) -- (2.5,-1.2) node [midway,fill=white] {$4$};

\draw (-3.5,-1.2) rectangle ++(1,-.5) node [pos=.5] {$\hat{u}_3 \!\oplus\! 1$};
\draw (-3,-2) circle [radius=.3] node {SC};
\draw (-1.5,-1.2) rectangle ++(1,-.5) node [pos=.5] {$\hat{u}_5 \!\oplus\! 1$};
\draw (-1,-2) circle [radius=.3] node {SC};
\draw (.5,-1.2) rectangle ++(1,-.5) node [pos=.5] {$\hat{u}_6 \!\oplus\! 1$};
\draw (1,-2) circle [radius=.3] node {SC};
\draw (2,-1.2) rectangle ++(1,-.5) node [pos=.5] {$\hat{u}_7 \!\oplus\! 1$};
\draw (2.5,-2) circle [radius=.3] node {SC};

\draw (-3,-2.3) -- (-3.5,-3.2) node [midway,fill=white] {$6$};
\draw (-3,-2.3) -- (-4.6,-3.2) node [midway,fill=white] {$5$};
\draw (-3,-2.3) -- (-2.4,-3.2) node [midway,fill=white] {$7$};

\draw (-1,-2.3) -- (-1,-3.2) node [midway,fill=white] {$8$};
\draw (-1,-2.3) -- (.1,-3.2) node [midway,fill=white] {$9$};

\draw (1,-2.3) -- (1.5,-3.2) node [midway,fill=white] {$10$};

\draw (-5.1,-3.2) rectangle ++(1,-.5) node [pos=.5] {$\hat{u}_5 \!\oplus\! 1$};
\draw (-4.6,-4) circle [radius=.3] node {SC};
\draw (-4,-3.2) rectangle ++(1,-.5) node [pos=.5] {$\hat{u}_6 \!\oplus\! 1$};
\draw (-3.5,-4) circle [radius=.3] node {SC};
\draw (-2.9,-3.2) rectangle ++(1,-.5) node [pos=.5] {$\hat{u}_7 \!\oplus\! 1$};
\draw (-2.4,-4) circle [radius=.3] node {SC};

\draw (-1.5,-3.2) rectangle ++(1,-.5) node [pos=.5] {$\hat{u}_6 \!\oplus\! 1$};
\draw (-1,-4) circle [radius=.3] node {SC};
\draw (-.4,-3.2) rectangle ++(1,-.5) node [pos=.5] {$\hat{u}_7 \!\oplus\! 1$};
\draw (.1,-4) circle [radius=.3] node {SC};

\draw (1,-3.2) rectangle ++(1,-.5) node [pos=.5] {$\hat{u}_7 \!\oplus\! 1$};
\draw (1.5,-4) circle [radius=.3] node {SC};

\draw (-4.6,-4.3) -- (-5,-5.2) node [midway,fill=white] {$11$};
\draw (-4.6,-4.3) -- (-3.9,-5.2) node [midway,fill=white] {$12$};

\draw (-3.5,-4.3) -- (-2.5,-5.2) node [midway,fill=white] {$13$};

\draw (-1,-4.3) -- (-1,-5.2) node [midway,fill=white] {$14$};

\draw (-5.5,-5.2) rectangle ++(1,-.5) node [pos=.5] {$\hat{u}_6 \!\oplus\! 1$};
\draw (-5,-6) circle [radius=.3] node {SC};
\draw (-4.4,-5.2) rectangle ++(1,-.5) node [pos=.5] {$\hat{u}_7 \!\oplus\! 1$};
\draw (-3.9,-6) circle [radius=.3] node {SC};

\draw (-3,-5.2) rectangle ++(1,-.5) node [pos=.5] {$\hat{u}_7 \!\oplus\! 1$};
\draw (-2.5,-6) circle [radius=.3] node {SC};

\draw (-1.5,-5.2) rectangle ++(1,-.5) node [pos=.5] {$\hat{u}_7 \!\oplus\! 1$};
\draw (-1,-6) circle [radius=.3] node {SC};

\draw (-5,-6.3) -- (-5,-7.2) node [midway,fill=white] {$15$};

\draw (-5.5,-7.2) rectangle ++(1,-.5) node [pos=.5] {$\hat{u}_7 \!\oplus\! 1$};
\draw (-5,-8) circle [radius=.3] node {SC};

\end{tikzpicture}
        \caption{BFS.}
        \label{fig:treeBFS}
    \end{subfigure}
    \caption{Bit-flipping tree for $\mathcal{RM}(1,3)$ with $\mathcal{F}=\{0,1,2,4\}$.}
    \label{fig:tree}
\end{figure*}
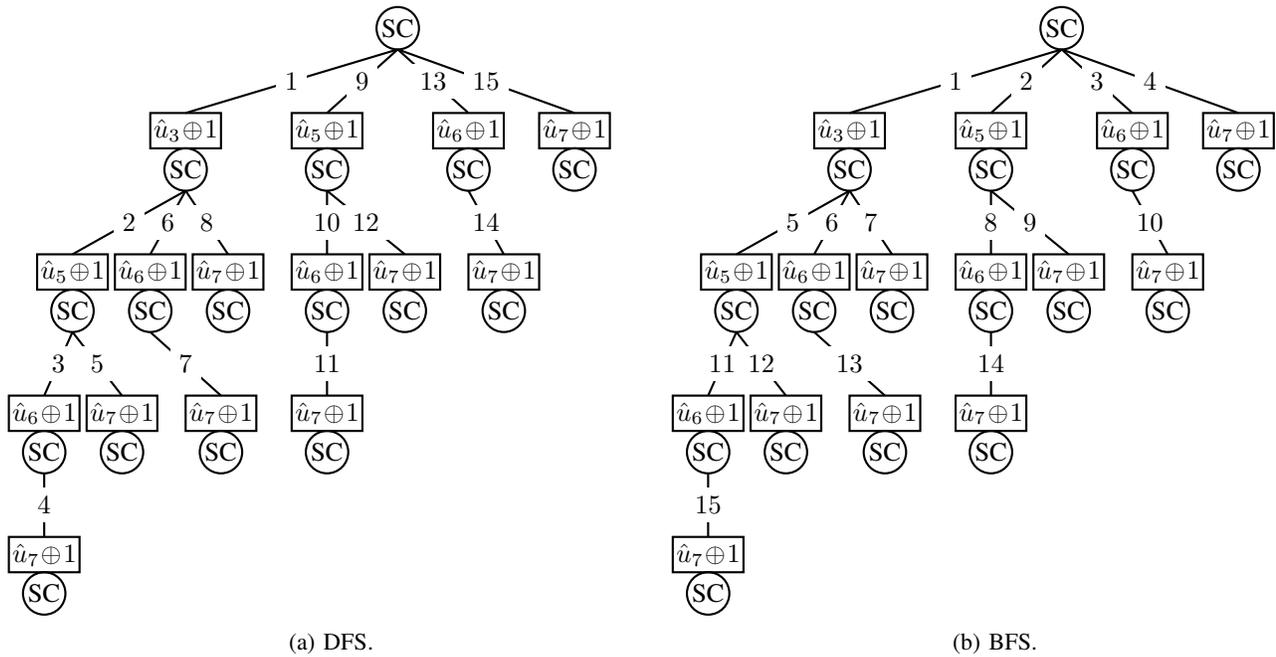

There are two main issues with SCOS decoding. First, the calculation of the score in (\ref{equ:score}) depends on estimating $p_j$ that is approximated either by massive Monte-Carlo simulations or by density evolution. This is especially cumbersome in rapidly changing environments where the score must be updated frequently. Second, the sorting of scores is a computationally intensive task, especially when the list size increases. The next section presents a tree search approach to tackle SCOS decoding issues.

\section{The Tree Search Approach}

\subsection{Proposed Algorithm}

The tree search approach relies on the generation of a bit-flipping tree as shown in Fig.~\ref{fig:tree}. Each node in the tree represents a SC decoding attempt after a bit-flipping is performed on the result of the SC decoder in its parent node. The proposed decoding scheme starts by an initial SC decoding attempt, which is represented as the root node in the tree. The corresponding candidate result is stored as $\hat{\bm{u}}_{\text{best}}$, and the corresponding path metric is calculated based on (\ref{equ:polar:PM_LLR}) and is stored as $\text{PM}_{\text{best}}$. The decoder then proceeds by traversing the bit-flipping tree until all the nodes are visited. Let the complexity of the proposed decoding scheme be the number of SC decoding attempts. It can be seen that the worst-case complexity of the proposed decoder is equivalent to the number of nodes in the bit-flipping tree, which is
\begin{equation}
    \mathcal{C}_1 = 2^K\text{.}
\end{equation}
To reduce the average complexity of the decoder, The tree needs to be pruned to limit the number of node visits. Therefore, an approach similar to the SCOS decoder in \cite{yuan_ML} is adopted. In particular, before each node visit, the path metric associated with the bit-flipping at that node is calculated and compared with $\text{PM}_{\text{best}}$. The node is visited only if the path metric is less than $\text{PM}_{\text{best}}$. Otherwise, that node and any node that is a descendent of that node is not visited. In the case of a node visit, the corresponding path metric is calculated based on (\ref{equ:polar:PM_LLR}) and is compared with $\text{PM}_{\text{best}}$. If the path metric is less than $\text{PM}_{\text{best}}$, then $\text{PM}_{\text{best}}$ is updated with the new path metric and the new ML candidate is stored as $\hat{\bm{u}}_{\text{best}}$.

To traverse the bit-flipping tree without calculating an additional metric and without any sorting operations, the natural order in the bit indices are considered and two tree traversal approaches are adopted. First, a DFS approach, in which the priority is given to the nested bit-flipping operations as shown in Fig.~\ref{fig:treeDFS} for $\mathcal{RM}(1,3)$. Second, a BFS approach, in which the priority is given to bit-flips at the same levels of the tree as shown in Fig.~\ref{fig:treeBFS} for $\mathcal{RM}(1,3)$. Note that the order in which the nodes are visited is indicated by the numbers on the edges of the bit-flipping tree. Algorithm~\ref{algo:RSC:DFS} and Algorithm~\ref{algo:RSC:BFS} summarize the functions $\texttt{TS-DFS}(\cdot)$ and $\texttt{TS-BFS}(\cdot)$, which represent the bit-flipping tree search based on the DFS and the BFS strategies, respectively.

In Algorithm~\ref{algo:RSC:DFS}, the inputs to the $\texttt{TS-DFS}(\cdot)$ function are $\bm{y}$, the set of bit-flipping indices $\mathcal{E}$, the ML candidate $\hat{\bm{u}}_{\text{best}}$, and the path metric associated with the ML candidate $\text{PM}_{\text{best}}$. Before the function is called, $\mathcal{E} = \emptyset$ and $\text{PM}_{\text{best}} = \infty$. The function starts by a round of SC decoding using the function $\texttt{SC}(\cdot)$, which takes $\bm{y}$ and $\mathcal{E}$ as inputs, and results in a candidate $\hat{\bm{u}}$ and its corresponding LLR values $\bm{\alpha}$. The path metric for $\hat{\bm{u}}$ is then calculated and stored in $\text{PM}_{\hat{\bm{u}}}$. The ML candidate is updated only if $\text{PM}_{\hat{\bm{u}}} < \text{PM}_{\text{best}}$. Then for any information bit whose index is larger than the maximum index in $\mathcal{E}$, denoted by $i^{\max}_\mathcal{E}$, a metric associated with the bit-flip at the $i$-th index is calculated and is stored in $\text{PM}_{\text{tmp}}$. A recursive call to $\texttt{TS-DFS}(\cdot)$ is conducted only if $\text{PM}_{\text{tmp}} < \text{PM}_{\text{best}}$. The $\texttt{TS-BFS}(\cdot)$ function in Algorithm~\ref{algo:RSC:BFS} depends on storing a set $\mathcal{S}$ of LLR values and bit-flipping positions as $\{\bm{\alpha},\mathcal{E}\}$. Since the tree search is conducted by giving priority to the node visits in the same level of the tree, all the candidates at the same level of tree with $\text{PM}_{\text{tmp}} < \text{PM}_{\text{best}}$ are stored in an auxiliary set $\mathcal{S}_{\text{next}}$. The set $\mathcal{S}$ is updated by $\mathcal{S}_{\text{next}}$ and the function runs until the set $\mathcal{S}$ becomes empty.

\begin{algorithm}[t]
	\DontPrintSemicolon
	\caption{$\texttt{TS-DFS}(\cdot)$}
	\label{algo:RSC:DFS}
	\SetKwInOut{Input}{Input}
	\SetKwInOut{Output}{Output}
	\SetKwInput{kwIn}{in}
	
	
	\SetKwFunction{FMain}{TS-DFS}
	\SetKwFunction{Sort}{Sort}
	\SetKwProg{Fn}{Function}{:}{}
	
	
	\Fn{\FMain{$\bm{y}, \mathcal{E}, \hat{\bm{u}}_\textup{best}, \textup{PM}_\textup{best}$}}{
		$\bm{\alpha}, \hat{\bm{u}}$ $\leftarrow$ \texttt{SC$(\bm{y}, \mathcal{E})$}\\
		
		$\textup{PM}_{\hat{\bm{u}}} \leftarrow \sum\limits_{j \in \mathcal{F}}\frac{1-\sgn(\alpha_j)}{2} \abs{\alpha_j} + \sum\limits_{j \in \mathcal{E}}\abs{\alpha_j}$
		
		\If{$\textup{PM}_{\hat{\bm{u}}} < \textup{PM}_\textup{best}$}{
			$\textup{PM}_\textup{best} \leftarrow \textup{PM}_{\hat{\bm{u}}}$\\
			$\hat{\bm{u}}_\textup{best} \leftarrow \hat{\bm{u}}$\\
		}
		
		\For{$i \in \{\mathcal{I}|i^{\max}_\mathcal{E} < i\}$} {
			$\textup{PM}_\textup{tmp} \leftarrow \sum\limits_{j \in \{\mathcal{F}| j<i\}}\frac{1-\sgn(\alpha_j)}{2} \abs{\alpha_j} + \sum\limits_{j \in \mathcal{E} \cup i}\abs{\alpha_j}$\\
			\If{$\textup{PM}_\textup{tmp} < \textup{PM}_\textup{best}$}{
				$\hat{\bm{u}}_\textup{best},\! \textup{PM}_\textup{best} \!\!\leftarrow\!\! \FMain(\bm{y},\mathcal{E}\cup i, \hat{\bm{u}}_\textup{best},\! \textup{PM}_\textup{best})$\\
			}
		}
		\Return{$\hat{\bm{u}}_\textup{best}, \textup{PM}_\textup{best}$}
	}	
\end{algorithm}

While the worst-case complexity of the DFS and the BFS schemes for searching the tree is exponential in $K$, simulation results in Section~\ref{sec:res} show that the average complexity of the proposed method is small, especially when the channel experiences low levels of noise.

\subsection{Complexity Considerations}

\subsubsection{Limiting the Tree Search}
\begin{figure*}[t]
    \centering
    \begin{subfigure}{.48\textwidth}
        \begin{tikzpicture}[scale=.94, thick]

\draw (0,0) circle [radius=.3] node {SC};

\draw (0,-.3) -- (-1,-1.2) node [midway,fill=white] {$5$};
\draw (0,-.3) -- (-3,-1.2) node [midway,fill=white] {$1$};
\draw (0,-.3) -- (1,-1.2) node [midway,fill=white] {$8$};
\draw (0,-.3) -- (2.5,-1.2) node [midway,fill=white] {$10$};

\draw (-3.5,-1.2) rectangle ++(1,-.5) node [pos=.5] {$\hat{u}_3 \!\oplus\! 1$};
\draw (-3,-2) circle [radius=.3] node {SC};
\draw (-1.5,-1.2) rectangle ++(1,-.5) node [pos=.5] {$\hat{u}_5 \!\oplus\! 1$};
\draw (-1,-2) circle [radius=.3] node {SC};
\draw (.5,-1.2) rectangle ++(1,-.5) node [pos=.5] {$\hat{u}_6 \!\oplus\! 1$};
\draw (1,-2) circle [radius=.3] node {SC};
\draw (2,-1.2) rectangle ++(1,-.5) node [pos=.5] {$\hat{u}_7 \!\oplus\! 1$};
\draw (2.5,-2) circle [radius=.3] node {SC};

\draw (-3,-2.3) -- (-3.5,-3.2) node [midway,fill=white] {$3$};
\draw (-3,-2.3) -- (-4.6,-3.2) node [midway,fill=white] {$2$};
\draw (-3,-2.3) -- (-2.4,-3.2) node [midway,fill=white] {$4$};

\draw (-1,-2.3) -- (-1,-3.2) node [midway,fill=white] {$6$};
\draw (-1,-2.3) -- (.1,-3.2) node [midway,fill=white] {$7$};

\draw (1,-2.3) -- (1.5,-3.2) node [midway,fill=white] {$9$};

\draw (-5.1,-3.2) rectangle ++(1,-.5) node [pos=.5] {$\hat{u}_5 \!\oplus\! 1$};
\draw (-4.6,-4) circle [radius=.3] node {SC};
\draw (-4,-3.2) rectangle ++(1,-.5) node [pos=.5] {$\hat{u}_6 \!\oplus\! 1$};
\draw (-3.5,-4) circle [radius=.3] node {SC};
\draw (-2.9,-3.2) rectangle ++(1,-.5) node [pos=.5] {$\hat{u}_7 \!\oplus\! 1$};
\draw (-2.4,-4) circle [radius=.3] node {SC};

\draw (-1.5,-3.2) rectangle ++(1,-.5) node [pos=.5] {$\hat{u}_6 \!\oplus\! 1$};
\draw (-1,-4) circle [radius=.3] node {SC};
\draw (-.4,-3.2) rectangle ++(1,-.5) node [pos=.5] {$\hat{u}_7 \!\oplus\! 1$};
\draw (.1,-4) circle [radius=.3] node {SC};

\draw (1,-3.2) rectangle ++(1,-.5) node [pos=.5] {$\hat{u}_7 \!\oplus\! 1$};
\draw (1.5,-4) circle [radius=.3] node {SC};

\end{tikzpicture}
        \caption{DFS.}
        \label{fig:treeDFSDepth}
    \end{subfigure}
    \begin{subfigure}{.48\textwidth}
        \begin{tikzpicture}[scale=.94, thick]

\draw (0,0) circle [radius=.3] node {SC};

\draw (0,-.3) -- (-1,-1.2) node [midway,fill=white] {$2$};
\draw (0,-.3) -- (-3,-1.2) node [midway,fill=white] {$1$};
\draw (0,-.3) -- (1,-1.2) node [midway,fill=white] {$3$};
\draw (0,-.3) -- (2.5,-1.2) node [midway,fill=white] {$4$};

\draw (-3.5,-1.2) rectangle ++(1,-.5) node [pos=.5] {$\hat{u}_3 \!\oplus\! 1$};
\draw (-3,-2) circle [radius=.3] node {SC};
\draw (-1.5,-1.2) rectangle ++(1,-.5) node [pos=.5] {$\hat{u}_5 \!\oplus\! 1$};
\draw (-1,-2) circle [radius=.3] node {SC};
\draw (.5,-1.2) rectangle ++(1,-.5) node [pos=.5] {$\hat{u}_6 \!\oplus\! 1$};
\draw (1,-2) circle [radius=.3] node {SC};
\draw (2,-1.2) rectangle ++(1,-.5) node [pos=.5] {$\hat{u}_7 \!\oplus\! 1$};
\draw (2.5,-2) circle [radius=.3] node {SC};

\draw (-3,-2.3) -- (-3.5,-3.2) node [midway,fill=white] {$6$};
\draw (-3,-2.3) -- (-4.6,-3.2) node [midway,fill=white] {$5$};
\draw (-3,-2.3) -- (-2.4,-3.2) node [midway,fill=white] {$7$};

\draw (-1,-2.3) -- (-1,-3.2) node [midway,fill=white] {$8$};
\draw (-1,-2.3) -- (.1,-3.2) node [midway,fill=white] {$9$};

\draw (1,-2.3) -- (1.5,-3.2) node [midway,fill=white] {$10$};

\draw (-5.1,-3.2) rectangle ++(1,-.5) node [pos=.5] {$\hat{u}_5 \!\oplus\! 1$};
\draw (-4.6,-4) circle [radius=.3] node {SC};
\draw (-4,-3.2) rectangle ++(1,-.5) node [pos=.5] {$\hat{u}_6 \!\oplus\! 1$};
\draw (-3.5,-4) circle [radius=.3] node {SC};
\draw (-2.9,-3.2) rectangle ++(1,-.5) node [pos=.5] {$\hat{u}_7 \!\oplus\! 1$};
\draw (-2.4,-4) circle [radius=.3] node {SC};

\draw (-1.5,-3.2) rectangle ++(1,-.5) node [pos=.5] {$\hat{u}_6 \!\oplus\! 1$};
\draw (-1,-4) circle [radius=.3] node {SC};
\draw (-.4,-3.2) rectangle ++(1,-.5) node [pos=.5] {$\hat{u}_7 \!\oplus\! 1$};
\draw (.1,-4) circle [radius=.3] node {SC};

\draw (1,-3.2) rectangle ++(1,-.5) node [pos=.5] {$\hat{u}_7 \!\oplus\! 1$};
\draw (1.5,-4) circle [radius=.3] node {SC};

\end{tikzpicture}
        \caption{BFS.}
        \label{fig:treeBFSDepth}
    \end{subfigure}
    \caption{Pruned bit-flipping tree for $\mathcal{RM}(1,3)$ with $\mathcal{F}=\{0,1,2,4\}$, considering a maximum tree depth of $\omega=2$.}
    \label{fig:treeDepth}
\end{figure*}
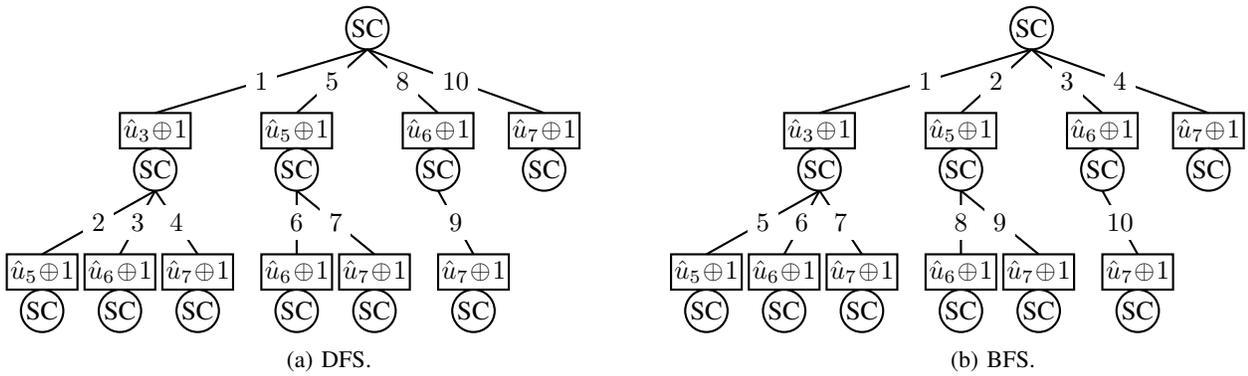

To reduce the number of node visits in the proposed decoding algorithm, the tree is pruned such that the search is only performed up to a specific depth in the tree. Let $\omega$ denote the maximum depth of the tree up to which the search is carried out and let $\texttt{TS-DFS-$\omega$}(\cdot)$ and $\texttt{TS-BFS-$\omega$}(\cdot)$ denote $\texttt{TS-DFS}(\cdot)$ and $\texttt{TS-BFS}(\cdot)$ functions that search the tree up to depth $\omega$, respectively. Fig.~\ref{fig:treeDepth} shows an example of this tree-pruning strategy for $\mathcal{RM}(1,3)$ with $\mathcal{F}=\{0,1,2,4\}$, considering the maximum depth of the tree is $\omega=2$, for both DFS (Fig.~\ref{fig:treeDFSDepth}) and BFS (Fig.~\ref{fig:treeBFSDepth}) schemes. Note that this tree-pruning strategy uses the fact that it is more likely to achieve the ML decoding performance by performing a small number of bit-flips \cite{DSCF}, which is equivalent to limiting the depth of the bit-flipping tree. The worst-case complexity of this scheme is calculated as
\begin{equation}
    \mathcal{C}_2 = \sum_{i=0}^{\omega} \binom{K}{i}\text{,}
\end{equation}
where $\mathcal{C}_2 < \mathcal{C}_1$ for $\omega < K$. It is worth noting that while this tree-pruning strategy significantly reduces the complexity of decoding, it does not guarantee the ML decoding performance. In addition, when $\omega=0$, the proposed decoding algorithm is equivalent to a single round of SC decoding, and when $\omega=1$, the DFS and the BFS strategies are equivalent.

\begin{algorithm}[t]
	\DontPrintSemicolon
	\caption{$\texttt{TS-BFS}(\cdot)$}
	\label{algo:RSC:BFS}
	\SetKwInOut{Input}{Input}
	\SetKwInOut{Output}{Output}
	\SetKwInput{kwIn}{in}
	
	
	\SetKwFunction{FMain}{TS-BFS}
	\SetKwFunction{Sort}{Sort}
	\SetKwProg{Fn}{Function}{:}{}
	
	\Fn{\FMain{$\bm{y}$}}{
	$\mathcal{E} \leftarrow \emptyset$\\
	$\bm{\alpha}, \hat{\bm{u}}_\textup{best}$ $\leftarrow$ \texttt{SC$(\bm{y})$}\\
	$\textup{PM}_{\textup{best}} \leftarrow \sum\limits_{j \in \mathcal{F}}\frac{1-\sgn(\alpha_j)}{2} \abs{\alpha_j}$
	
	$\mathcal{S} \leftarrow \{\{\bm{\alpha}, \mathcal{E}\}\}$
	
	\While{$\abs{\mathcal{S}}>0$}
	{
		$\mathcal{S}_\text{next} \leftarrow \emptyset$\\
		\For{$\{\bm{\alpha},\mathcal{E}\} \in \mathcal{S}$}
		{
			\For{$i \in \{\mathcal{I}|i^{\max}_\mathcal{E} < i\}$}
			{
				$\textup{PM}_\textup{tmp} \leftarrow \sum\limits_{j \in \{\mathcal{F}| j<i\}}\frac{1-\sgn(\alpha_j)}{2} \abs{\alpha_j} + \sum\limits_{j \in \mathcal{E} \cup i}\abs{\alpha_j}$\\
				\If{$\textup{PM}_\textup{tmp} < \textup{PM}_\textup{best}$}{
					$\bm{\alpha}_\textup{tmp}, \hat{\bm{u}}_\textup{tmp}$ $\leftarrow$ \texttt{SC$(\bm{y}, \mathcal{E} \cup i)$}\\
					
					$\textup{PM}_{\hat{\bm{u}}_\textup{tmp}} \leftarrow \sum\limits_{j \in \mathcal{F}}\frac{1-\sgn(\alpha_{\textup{tmp}_j})}{2} \abs{\alpha_{\textup{tmp}_j}} + \sum\limits_{j \in \mathcal{E} \cup i}\abs{\alpha_{\textup{tmp}_j}}$
					
					\If{$\textup{PM}_{\hat{\bm{u}}} < \textup{PM}_\textup{best}$}{
						$\textup{PM}_\textup{best} \leftarrow \textup{PM}_{\hat{\bm{u}}_\textup{tmp}}$\\
						$\hat{\bm{u}}_\textup{best} \leftarrow \hat{\bm{u}}_\textup{tmp}$\\
					}
					
					\If{$\textup{PM}_\textup{tmp} < \textup{PM}_\textup{best}$}
					{
						$\mathcal{S}_\textup{next} \leftarrow \mathcal{S}_\textup{next} \cup \{\bm{\alpha}_\textup{tmp}, \mathcal{E} \cup {i}\}$
					}
				}
			}
		}
		$\mathcal{S} \leftarrow \mathcal{S}_\textup{next}$
	}
	
	\Return $\hat{\bm{u}}_\textup{best}$
	}
\end{algorithm}

\subsubsection{Eliminating Redundant Node Visits}
\begin{figure}[t]
    \centering
    \begin{tikzpicture}[scale=.94, thick]

\draw (0,0) circle [radius=.3] node {SC};

\draw (0,-.3) -- (-3,-1.2) node [midway,fill=white] {$1$};

\draw (-3.5,-1.2) rectangle ++(1,-.5) node [pos=.5] {$\hat{u}_3 \!\oplus\! 1$};
\draw (-3,-2) circle [radius=.3] node {SC};

\end{tikzpicture}
    \caption{Pruned bit-flipping tree for $\mathcal{RM}(1,3)$ with $\mathcal{F}=\{0,1,2,4\}$. Since there is only one information bit ($u_3$) before the last frozen bit ($u_4$), a maximum of two rounds of SC decoding results in ML decoding performance.}
    \label{fig:treeLast}
\end{figure}
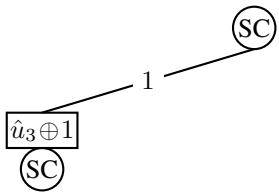

The path metric calculation in (\ref{equ:polar:PM_LLR}) significantly reduces the worst-case complexity by only considering the information bits that are located before the last frozen bit in the code. This is due to the fact that the path metrics in SC decoding (\ref{equ:polar:PM_LLR}) can only increase if the estimation of a bit does not follow the sign of its LLR value. Since the SC decoding algorithm is such that the estimation of each bit follows the sign of its LLR value, only frozen bits with fixed values can change the path metric in (\ref{equ:polar:PM_LLR}). The bit-by-bit schedule of SC decoding, thus, makes the estimation of any information bit after the last frozen bit in the code redundant \cite{Ali_PSCL_TCOM}. Let $K_l$ denote the number of information bits after the last frozen bit. The worst-case complexity of the proposed method is
\begin{equation}
    \mathcal{C}_3 = 2^{K-K_l}\text{.}
\end{equation}
Fig.~\ref{fig:treeLast} shows the pruned bit-flipping tree for $\mathcal{RM}(1,3)$ with $\mathcal{F}=\{0,1,2,4\}$. It can be seen that since bits $u_5$, $u_6$, and $u_7$ are located after the last frozen bit ($u_4$), the ML decoding result can be found by a maximum of two rounds of SC decoding. Note that since this optimization scheme can guarantee ML decoding performance, it is used in this paper in all the proposed methods.

\subsubsection{Ordering the Node Visits}
To reduce the average number of node visits in the proposed tree search approach, a metric is assigned to each set of bit-flips $\mathcal{E}$ and the node that results in the smallest metric is given priority at a specific level of the bit-flipping tree. This paper uses the metric developed in \cite{doan2020neural} as
\begin{equation} \label{equ:sortMetric}
    M_i = \smashoperator[lr]{\sum_{j \in \mathcal{E}\cup i}} \abs{\alpha_j} + \smashoperator[lr]{\sum_{j \in \{\mathcal{I}|j \leq i\}}} \max(0,\beta - \abs{\alpha_j})\text{,}
\end{equation}
where $\beta$ is a parameter that is empirically selected to be $0.8$. This approach ensures the bit-flipping set $\mathcal{E}$ that more likely caused the error is identified in the earlier stages of the tree search method, which in turn reduces the number of node visits. Let $\texttt{TS-DFS-O}(\cdot)$ and $\texttt{TS-BFS-O}(\cdot)$ denote $\texttt{TS-DFS}(\cdot)$ and $\texttt{TS-BFS}(\cdot)$ functions that use (\ref{equ:sortMetric}) to order the node visits, and $\texttt{TS-DFS-O-$\omega$}(\cdot)$ and $\texttt{TS-BFS-O-$\omega$}(\cdot)$ denote $\texttt{TS-DFS-$\omega$}(\cdot)$ and $\texttt{TS-BFS-$\omega$}(\cdot)$ functions that use (\ref{equ:sortMetric}) to order the node visits. Note that while this approach on average reduces the number of node visits, a maximum of $K$ metric values in (\ref{equ:sortMetric}) need to be sorted at each level of the bit-flipping tree.

\section{Numerical Results}\label{sec:res}

\begin{figure*}[t]
	\centering
	\begin{tikzpicture}
\pgfplotsset{
	label style = {font=\fontsize{9pt}{7.2}\selectfont},
	tick label style = {font=\fontsize{7pt}{7.2}\selectfont}
}

\begin{axis}[
scale = 1,
ymode=log,
xlabel={$E_b/N_0$ [\text{dB}]},
ytick={1e-6, 1e-5,1e-4,1e-3,1e-2,1e-1,1e-0},
xtick={2,3,4},
xmax=4.2,
ymax = 0.4,
ylabel={FER},
grid=both,
ymajorgrids=true,
xmajorgrids=true,
grid style=solid,
width=0.35\linewidth, height=6cm,
thick,
mark size=2.25,
legend cell align=left,
legend style={
	column sep= 2mm,
	font=\fontsize{7pt}{7.2}\selectfont,
},
legend to name=perf-legend,
legend columns=6,
]

\addplot[
	mark=square,
	thick,
	red,
	mark size=3,
	mark options={solid},
]
table {
	2	1.39E-01
	2.5	6.07E-02
	3.00E+00	2.19E-02
	3.5	5.87E-03
	4	1.27E-03	
};
\addlegendentry{$\texttt{TS-DFS-$1$}(\cdot)$}

\addplot[
	mark=triangle,
	thick,
	red,
	mark size=3,
	mark options={solid},
]
table {
	2	5.21E-02
	2.5	1.67E-02
	3.00E+00	4.35E-03
	3.5	8.36E-04
	4	9.97E-05
};
\addlegendentry{$\texttt{TS-DFS-$2$}(\cdot)$}

\addplot[
	mark=o,
	thick,
	red,
	mark size=3,
	mark options={solid},
]
table {
	2	2.58E-02
	2.5	6.82E-03
	3.00E+00	1.39E-03
	3.5	1.68E-04
	4	1.86E-05
};
\addlegendentry{$\texttt{TS-DFS-$3$}(\cdot)$}

\addplot[
	mark=x,
	thick,
	red,
	mark size=3,
	mark options={solid},
]
table {
	2	1.89E-02
	2.5	4.56E-03
	3.00E+00	7.52E-04
	3.5	9.27E-05
	4.00E+00	1.22E-05
};
\addlegendentry{$\texttt{TS-DFS-$4$}(\cdot)$}

\addplot[
	mark=pentagon,
	thick,
	red,
	mark size=3,
	mark options={solid},
]
table {
	2	2
};
\addlegendentry{$\texttt{TS-DFS-$5$}(\cdot)$}

\addplot[
	black,
	thick,
	dashdotted,
	mark options={solid},
]
table {
	2.00E+00	1.69E-02
	2.50E+00	4.09E-03
	3.00E+00	6.51E-04
	3.5	8.68E-05
	4	1.17E-05
};
\addlegendentry{ML (lower bound)}

\addplot[
	mark=square,
	thick,
	blue,
	mark size=3,
	dotted,
	mark options={solid},
]
table {
	2	1.39E-01
	2.5	6.07E-02
	3.00E+00	2.19E-02
	3.5	5.87E-03
	4	1.27E-03	
};
\addlegendentry{$\texttt{TS-BFS-$1$}(\cdot)$}

\addplot[
	mark=triangle,
	thick,
	blue,
	mark size=3,
	dotted,
	mark options={solid},
]
table {
	2	5.21E-02
	2.5	1.67E-02
	3.00E+00	4.35E-03
	3.5	8.36E-04
	4	9.97E-05
};
\addlegendentry{$\texttt{TS-BFS-$2$}(\cdot)$}

\addplot[
	mark=o,
	thick,
	blue,
	mark size=3,
	dotted,
	mark options={solid},
]
table {
	2	2.58E-02
	2.5	6.82E-03
	3.00E+00	1.39E-03
	3.5	1.68E-04
	4	1.86E-05
};
\addlegendentry{$\texttt{TS-BFS-$3$}(\cdot)$}

\addplot[
	mark=x,
	thick,
	blue,
	mark size=3,
	dotted,
	mark options={solid},
]
table {
	2	1.89E-02
	2.5	4.56E-03
	3.00E+00	7.52E-04
	3.5	9.27E-05
	4.00E+00	1.22E-05
};
\addlegendentry{$\texttt{TS-BFS-$4$}(\cdot)$}

\addplot[
	mark=pentagon,
	thick,
	blue,
	mark size=3,
	dotted,
	mark options={solid},
]
table {
	2	2
};
\addlegendentry{$\texttt{TS-BFS-$5$}(\cdot)$}



\node[anchor=north east, fill=white] at (rel axis cs:1,1) {\footnotesize{$\mathcal{RM}(3,7)$}};
\end{axis}
\end{tikzpicture}
	\begin{tikzpicture}
	\pgfplotsset{
		label style = {font=\fontsize{9pt}{7.2}\selectfont},
		tick label style = {font=\fontsize{7pt}{7.2}\selectfont}
	}
	
	\begin{axis}[
		scale = 1,
		ymode=log,
		xlabel={$E_b/N_0$ [\text{dB}]},
		ytick={1e-6, 1e-5,1e-4,1e-3,1e-2,1e-1,1e-0},
		ymax=1,
		ylabel={},
		grid=both,
		ymajorgrids=true,
		xmajorgrids=true,
		grid style=solid,
		width=0.35\linewidth, height=6cm,
		thick,
		mark size=2.25,
		legend cell align=left,
		legend style={
			column sep= 2mm,
			font=\fontsize{7pt}{7.2}\selectfont,
		},
		]
		
		\addplot[
		mark=square,
		thick,
		red,
		mark size=3,
		mark options={solid},
		]
		table {
			2	4.57E-01
			2.5	2.64E-01
			3	1.21E-01
			3.5	4.37E-02
			4.00E+00	1.16E-02
			4.5	2.26E-03
			5.00E+00	3.47E-04
		};
		
		\addplot[
		mark=triangle,
		thick,
		red,
		mark size=3,
		mark options={solid},
		]
		table {
			2	3.44E-01
			2.5	1.68E-01
			3	6.21E-02
			3.5	1.75E-02
			4.00E+00	3.39E-03
			4.5	5.75E-04
			5.00E+00	6.60E-05			
		};
		
		\addplot[
		mark=o,
		thick,
		red,
		mark size=3,
		mark options={solid},
		]
		table {
			2	3.06E-01
			2.5	1.40E-01
			3	4.79E-02
			3.5	1.23E-02
			4.00E+00	2.33E-03
			4.5	3.74E-04
			5.00E+00	4.87E-05
		};
		
		
		\addplot[
		mark=square,
		thick,
		blue,
		mark size=3,
		dotted,
		mark options={solid},
		]
		table {
			2	4.57E-01
			2.5	2.64E-01
			3	1.21E-01
			3.5	4.37E-02
			4.00E+00	1.16E-02
			4.5	2.26E-03
			5.00E+00	3.47E-04
		};
		
		\addplot[
		mark=triangle,
		thick,
		blue,
		mark size=3,
		dotted,
		mark options={solid},
		]
		table {
			2	3.44E-01
			2.5	1.68E-01
			3	6.21E-02
			3.5	1.75E-02
			4.00E+00	3.39E-03
			4.5	5.75E-04
			5.00E+00	6.60E-05
		};
		
		\addplot[
		mark=o,
		thick,
		blue,
		mark size=3,
		dotted,
		mark options={solid},
		]
		table {
			2	3.06E-01
			2.5	1.40E-01
			3	4.79E-02
			3.5	1.23E-02
			4.00E+00	2.33E-03
			4.5	3.74E-04
			5.00E+00	4.87E-05
		};
		
		
		\addplot[
		black,
		thick,
		dashdotted,
		mark options={solid},
		]
		table {
			2	2.94E-01
			2.5	1.33E-01
			3	4.48E-02
			3.5	1.16E-02
			4	2.18E-03
			4.5	3.63E-04
			5	4.58E-05
		};
		
		\node[anchor=north east, fill=white] at (rel axis cs:1,1) {\footnotesize{$\mathcal{RM}(4,7)$}};
	\end{axis}
\end{tikzpicture}
	\begin{tikzpicture}
	\pgfplotsset{
		label style = {font=\fontsize{9pt}{7.2}\selectfont},
		tick label style = {font=\fontsize{7pt}{7.2}\selectfont}
	}
	
	\begin{axis}[
		scale = 1,
		ymode=log,
		xlabel={$E_b/N_0$ [\text{dB}]},
		ytick={1e-6, 1e-5,1e-4,1e-3,1e-2,1e-1,1e-0},
		ylabel={},
		grid=both,
		ymajorgrids=true,
		xmajorgrids=true,
		grid style=solid,
		width=0.35\linewidth, height=6cm,
		thick,
		mark size=2.25,
		legend cell align=left,
		legend style={
			column sep= 2mm,
			font=\fontsize{7pt}{7.2}\selectfont,
		},
		]
		
		\addplot[
		mark=square,
		thick,
		red,
		mark size=3,
		mark options={solid},
		]
		table {	
			2	1.29E-01
			2.5	6.79E-02
			3	3.18E-02
			3.5	1.23E-02
			4	3.68E-03
			4.5	1.03E-03
			5.00E+00	2.12E-04
		};
		
		\addplot[
		mark=triangle,
		thick,
		red,
		mark size=3,
		mark options={solid},
		]
		table {
			2	3.26E-02
			2.5	1.28E-02
			3	4.35E-03
			3.5	1.15E-03
			4	2.35E-04
			4.5	3.73E-05
		};
		
		\addplot[
		mark=o,
		thick,
		red,
		mark size=3,
		mark options={solid},
		]
		table {
			2	7.98E-03
			2.5	2.32E-03
			3	5.49E-04
			3.5	1.14E-04
			4.00E+00	1.49E-05
		};
		
		\addplot[
			mark=x,
			thick,
			red,
			mark size=3,
			mark options={solid},
		]
		table {
			2	2.52E-03
			2.5	5.62E-04
			3	9.06E-05
			3.50E+00	1.66E-05
		};
		
		\addplot[
			mark=pentagon,
			thick,
			red,
			mark size=3,
			mark options={solid},
		]
		table {
			2	1.45E-03
			2.5	2.33E-04
			3.00E+00	2.89E-05
		};
		
		\addplot[
		mark=square,
		thick,
		blue,
		mark size=3,
		dotted,
		mark options={solid},
		]
		table {	
			2	1.29E-01
			2.5	6.79E-02
			3	3.18E-02
			3.5	1.23E-02
			4	3.68E-03
			4.5	1.03E-03
			5.00E+00	2.12E-04
		};
		
		\addplot[
		mark=triangle,
		thick,
		blue,
		mark size=3,
		dotted,
		mark options={solid},
		]
		table {
			2	3.26E-02
			2.5	1.28E-02
			3	4.35E-03
			3.5	1.15E-03
			4	2.35E-04
			4.5	3.73E-05
		};
		
		\addplot[
		mark=o,
		thick,
		blue,
		mark size=3,
		dotted,
		mark options={solid},
		]
		table {
			2	7.98E-03
			2.5	2.32E-03
			3	5.49E-04
			3.5	1.14E-04
			4.00E+00	1.49E-05
		};
		
		\addplot[
			mark=x,
			thick,
			blue,
			mark size=3,
    		dotted,
			mark options={solid},
		]
		table {
			2	2.52E-03
			2.5	5.62E-04
			3	9.06E-05
			3.50E+00	1.66E-05
		};
		
		\addplot[
			mark=pentagon,
			thick,
			blue,
			mark size=3,
    		dotted,
			mark options={solid},
		]
		table {
			2	1.45E-03
			2.5	2.33E-04
			3.00E+00	2.89E-05
		};

		\addplot[
		black,
		thick,
		dashdotted,
		mark options={solid},
		]
		table {
			2	1.23E-03
			2.5	2.14E-04
			3	2.44E-05
		};
		
		\node[anchor=north east, fill=white] at (rel axis cs:1,1) {\footnotesize{$\mathcal{RM}(2,8)$}};
	\end{axis}
\end{tikzpicture}
	\ref{perf-legend}\\
	\caption{Error-correction performance of $\mathcal{RM}(3,7)$, $\mathcal{RM}(4,7)$, and $\mathcal{RM}(2,8)$ under $\texttt{TS-DFS-$\omega$}(\cdot)$ and $\texttt{TS-BFS-$\omega$}(\cdot)$ for different values of $\omega$.}
	\label{fig:fer}
\end{figure*}
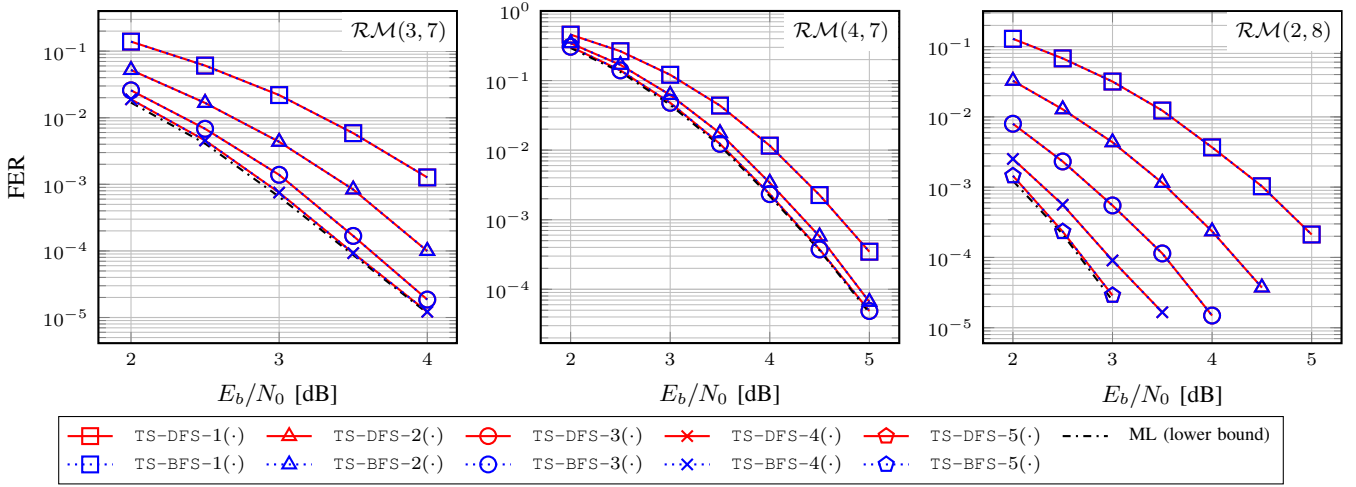

This section evaluates the frame error rate (FER) performance and the complexity of the proposed tree search methods and compares them with the SCOS decoder in \cite{yuan_ML}. The functions $\texttt{TS-DFS}(\cdot)$, $\texttt{TS-BFS}(\cdot)$, $\texttt{TS-DFS-O}(\cdot)$, and $\texttt{TS-BFS-O}(\cdot)$ achieve the ML decoding performance for any RM code. Fig.~\ref{fig:fer} shows the FER performance of $\mathcal{RM}(3,7)$, $\mathcal{RM}(4,7)$, and $\mathcal{RM}(2,8)$ using the functions $\texttt{TS-DFS-$\omega$}(\cdot)$ and $\texttt{TS-BFS-$\omega$}(\cdot)$ for different values of $\omega$. It can be seen that while both $\texttt{TS-DFS-$\omega$}(\cdot)$ and $\texttt{TS-BFS-$\omega$}(\cdot)$ decoders provide the same FER performance at the same value of $\omega$, selecting $\omega=4$ for $\mathcal{RM}(3,7)$, $\omega=3$ for $\mathcal{RM}(4,7)$, and $\omega=5$ for $\mathcal{RM}(2,8)$ results in an FER performance that is very close to the lower bound of ML decoding performance.

\begin{figure*}[t]
	\centering
	\begin{tikzpicture}
\pgfplotsset{
	label style = {font=\fontsize{9pt}{7.2}\selectfont},
	tick label style = {font=\fontsize{7pt}{7.2}\selectfont}
}

\begin{axis}[
scale = 1,
ymode=log,
xlabel={$E_b/N_0$ [\text{dB}]},
ylabel={$\mathcal{C}_\text{avg}$},
grid=both,
ymajorgrids=true,
xmajorgrids=true,
grid style=solid,
width=0.35\linewidth, height=6cm,
thick,
mark size=2.25,
legend cell align=left,
legend style={
	column sep= 2mm,
	font=\fontsize{7pt}{7.2}\selectfont,
},
legend to name=comp-legend,
legend columns=5,
]

\addplot[
mark=square,
thick,
red,
mark size=3,
mark options={solid},
]
table {
	2	1.08E+02
	2.5	4.76E+01
	3.00E+00	1.83E+01
	3.5	6.54E+00
	4.00E+00	2.41E+00
};
\addlegendentry{$\texttt{TS-DFS-$\omega$}(\cdot)$}

\addplot[
	mark=triangle,
	thick,
	red,
	mark size=3,
	solid,
	mark options={solid},
]
table {
	2.00E+00	4.27E+02
	2.50E+00	1.62E+02
	3.00E+00	5.22E+01
	3.5	1.50E+01
	4	4.03E+00
};
\addlegendentry{$\texttt{TS-DFS}(\cdot)$}

\addplot[
	mark=o,
	thick,
	blue,
	mark size=3,
	mark options={solid},
]
table {
	2	4.71E+01
	2.5	1.70E+01
	3.00E+00	6.00E+00
	3.5	2.44E+00
	4.00E+00	1.38E+00
};
\addlegendentry{$\texttt{TS-BFS-$\omega$}(\cdot)$}

\addplot[
	mark=pentagon,
	thick,
	blue,
	mark size=3,
	mark options={solid},
]
table {
	2	6.86E+01
	2.5	2.06E+01
	3	6.43E+00
	3.5	2.48E+00
	4.00E+00	1.39E+00
};
\addlegendentry{$\texttt{TS-BFS}(\cdot)$}

\addplot[
mark=diamond,
thick,
black,
mark size=3,
mark options={solid},
]
table {
	2	9.97E+01
	2.5	3.27E+01
	3	1.07E+01
	3.5	3.87E+00
	4.00E+00	1.79E+00
};
\addlegendentry{SCOS \cite{yuan_ML}}

		\addplot[
		mark=square,
		thick,
		red,
		mark size=3,
		dashed,
		mark options={solid},
		]
		table {
			2	8.89E+01
			2.5	3.74E+01
			3.00E+00	1.43E+01
			3.5	5.14E+00
			4.00E+00	2.11E+00
		};
		\addlegendentry{$\texttt{TS-DFS-O-$\omega$}(\cdot)$}

		\addplot[
			mark=triangle,
			thick,
			red,
			mark size=3,
			dashed,
			mark options={solid},
		]
		table {
			2	2.11E+02
			2.5	7.73E+01
			3	2.49E+01
			3.5	7.48E+00
			4	2.52E+00
		};
		\addlegendentry{$\texttt{TS-DFS-O}(\cdot)$}
		
		\addplot[
		mark=o,
		thick,
		blue,
		mark size=3,
		dashed,
		mark options={solid},
		]
		table {
			2	4.68E+01
			2.5	1.70E+01
			3	6.07E+00
			3.5	2.49E+00
			4.00E+00	1.41E+00
		};
		\addlegendentry{$\texttt{TS-BFS-O-$\omega$}(\cdot)$}
		
		\addplot[
			mark=pentagon,
			thick,
			blue,
			mark size=3,
			dashed,
			mark options={solid},
		]
		table {
			2	6.80E+01
			2.5	2.03E+01
			3	6.33E+00
			3.5	2.44E+00
			4.00E+00	1.38E+00
		};
		\addlegendentry{$\texttt{TS-BFS-O}(\cdot)$}

\node[anchor=north east, fill=white] at (rel axis cs:1,1) {\footnotesize{$\mathcal{RM}(3,7)$}};
\end{axis}
\end{tikzpicture}
	\begin{tikzpicture}
	\pgfplotsset{
		label style = {font=\fontsize{9pt}{7.2}\selectfont},
		tick label style = {font=\fontsize{7pt}{7.2}\selectfont}
	}
	
	\begin{axis}[
		scale = 1,
		ymode=log,
		xlabel={$E_b/N_0$ [\text{dB}]},
		ylabel={},
		grid=both,
		ymajorgrids=true,
		xmajorgrids=true,
		grid style=solid,
		width=0.35\linewidth, height=6cm,
		thick,
		mark size=2.25,
		legend cell align=left,
		legend style={
			column sep= 2mm,
			font=\fontsize{7pt}{7.2}\selectfont,
		},
		]
			
		\addplot[
			mark=square,
			thick,
			red,
			mark size=3,
			mark options={solid},
		]
		table {
			2	6.35E+01
			2.5	3.65E+01
			3.00E+00	1.77E+01
			3.50E+00	7.48E+00
			4.00E+00	3.07E+00
			4.5	1.51E+00
			5.00E+00	1.10E+00
		};
		
		\addplot[
		mark=triangle,
		thick,
		red,
		mark size=3,
		mark options={solid},
		]
		table {
			2	1.88E+02
			2.5	1.03E+02
			3	4.63E+01
			3.5	1.73E+01
			4	5.80E+00
			4.5	2.10E+00
			5	1.19E+00
		};
				
		\addplot[
		mark=o,
		thick,
		blue,
		mark size=3,
		mark options={solid},
		]
		table {
			2	4.76E+01
			2.5	2.45E+01
			3.00E+00	1.05E+01
			3.50E+00	4.06E+00
			4.00E+00	1.82E+00
			4.5	1.18E+00
			5.00E+00	1.03E+00
		};
		
		\addplot[
		mark=pentagon,
		thick,
		blue,
		mark size=3,
		mark options={solid},
		]
		table {
			2	7.40E+01
			2.5	3.44E+01
			3	1.31E+01
			3.5	4.54E+00
			4	1.88E+00
			4.5	1.18E+00
			5	1.03E+00
		};
		
		\addplot[
		mark=diamond,
		thick,
		black,
		mark size=3,
		mark options={solid},
		]
		table {
			2	1.14E+02
			2.5	5.66E+01
			3	2.34E+01
			3.5	8.63E+00
			4	3.21E+00
			4.5	1.54E+00
			5	1.11E+00
		};
		
		\addplot[
			mark=square,
			thick,
			red,
			mark size=3,
			dashed,
			mark options={solid},
		]
		table {
			2	5.75E+01
			2.5	3.17E+01
			3.00E+00	1.47E+01
			3.50E+00	5.99E+00
			4.00E+00	2.51E+00
			4.5	1.36E+00
			5.00E+00	1.07E+00
		};
		
		\addplot[
		mark=triangle,
		thick,
		red,
		mark size=3,
		dashed,
		mark options={solid},
		]
		table {
			2	1.21E+02
			2.5	6.17E+01
			3	2.60E+01
			3.5	9.35E+00
			4	3.35E+00
			4.5	1.51E+00
			5	1.09E+00
		};
		
		\addplot[
		mark=o,
		thick,
		blue,
		mark size=3,
		dashed,
		mark options={solid},
		]
		table {
			2	4.65E+01
			2.5	2.40E+01
			3.00E+00	1.03E+01
			3.50E+00	4.04E+00
			4.00E+00	1.83E+00
			4.5	1.19E+00
			5.00E+00	1.03E+00
		};
		
		\addplot[
		mark=pentagon,
		thick,
		blue,
		mark size=3,
		dashed,
		mark options={solid},
		]
		table {
			2	7.28E+01
			2.5	3.36E+01
			3	1.26E+01
			3.5	4.34E+00
			4	1.81E+00
			4.5	1.16E+00
			5	1.03E+00
		};
		
		\node[anchor=north east, fill=white] at (rel axis cs:1,1) {\footnotesize{$\mathcal{RM}(4,7)$}};
	\end{axis}
\end{tikzpicture}
	\begin{tikzpicture}
	\pgfplotsset{
		label style = {font=\fontsize{9pt}{7.2}\selectfont},
		tick label style = {font=\fontsize{7pt}{7.2}\selectfont}
	}
	
	\begin{axis}[
		scale = 1,
		ymode=log,
		xlabel={$E_b/N_0$ [\text{dB}]},
		ytick={1e0,1e1,1e2,1e3},
		ylabel={},
		grid=both,
		ymajorgrids=true,
		xmajorgrids=true,
		grid style=solid,
		width=0.35\linewidth, height=6cm,
		thick,
		mark size=2.25,
		legend cell align=left,
		legend style={
			column sep= 2mm,
			font=\fontsize{7pt}{7.2}\selectfont,
		},
		]
		
		\addplot[
			mark=square,
			thick,
			red,
			mark size=3,
			mark options={solid},
		]
		table {
			2	4.4485E+02
			2.5	2.5725E+02
			3	1.3492E+02
		};
		
		\addplot[
		mark=triangle,
		thick,
		red,
		mark size=3,
		mark options={solid},
		]
		table {
			2	1.46E+03
			2.5	7.41E+02
			3	3.37E+02
		};
		
		\addplot[
		mark=o,
		thick,
		blue,
		mark size=3,
		mark options={solid},
		]
		table {
			2.0	1.90694E+02
			2.5	1.08480E+02
			3.0	5.72108E+01
		};
		
		\addplot[
			mark=pentagon,
			thick,
			blue,
			mark size=3,
			mark options={solid},
		]
		table {
			2	2.42E+02
			2.5	1.27E+02
			3	6.32E+01
		};
		
		\addplot[
		mark=diamond,
		thick,
		black,
		mark size=3,
		mark options={solid},
		]
		table {
			2	2.964E+02
			2.5	1.506E+02
			3	7.286E+01
		};
		
		\addplot[
		mark=square,
		thick,
		red,
		mark size=3,
		dashed,
		mark options={solid},
		]
		table {
			2	3.76348E+02
			2.5	2.18334E+02
			3	1.16117E+02	
		};
		
		\addplot[
		mark=triangle,
		thick,
		red,
		mark size=3,
		dashed,
		mark options={solid},
		]
		table {
			2	7.521E+02
			2.5	3.821E+02
			3	1.806E+02
		};
		
		\addplot[
			mark=o,
			thick,
			blue,
			mark size=3,
			dashed,
			mark options={solid},
		]
		table {
			2	1.91E+02
			2.5	1.08E+02
			3.00E+00	5.72E+01
		};
		
		\addplot[
		mark=pentagon,
		thick,
		blue,
		mark size=3,
		dashed,
		mark options={solid},
		]
		table {
			2	2.41E+02
			2.5	1.26E+02
			3	6.31E+01
		};
		
		\node[anchor=north east, fill=white] at (rel axis cs:1,1) {\footnotesize{$\mathcal{RM}(2,8)$}};
	\end{axis}
\end{tikzpicture}
	\ref{comp-legend}\\
	\caption{Average decoding complexity in terms of the number of SC decoding attempts for $\mathcal{RM}(3,7)$, $\mathcal{RM}(4,7)$, and $\mathcal{RM}(2,8)$ under different tree search methods in this paper in comparison with the SCOS decoder of \cite{yuan_ML}. Note that $\omega=4$ for $\mathcal{RM}(3,7)$, $\omega=3$ for $\mathcal{RM}(4,7)$, and $\omega=5$ for $\mathcal{RM}(2,8)$.}
	\label{fig:comp}
\end{figure*}
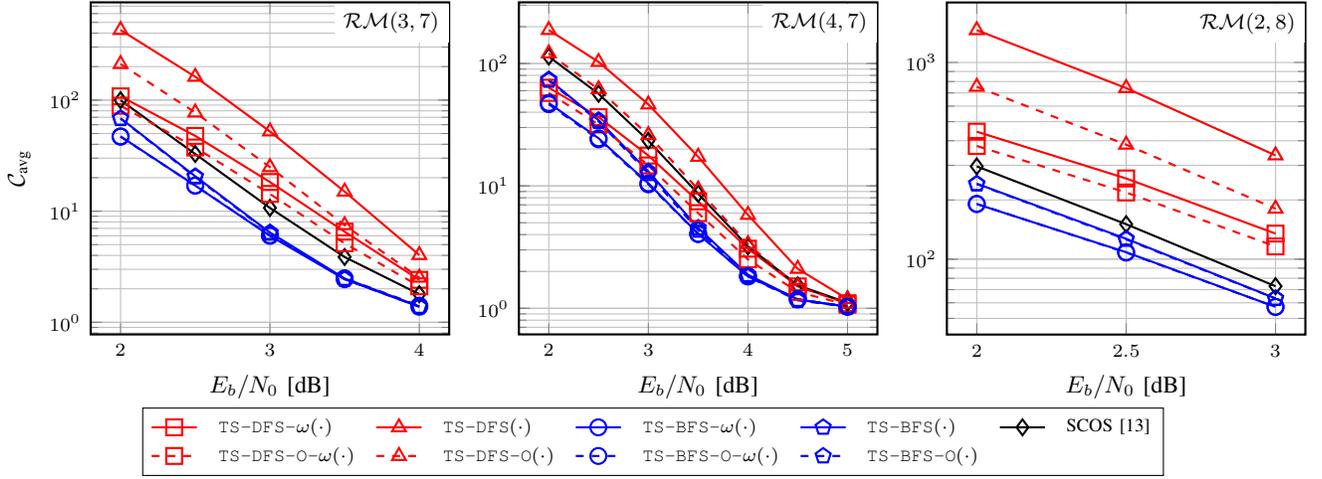

Fig.~\ref{fig:comp} shows the average decoding complexity $\mathcal{C}_{\text{avg}}$ in terms of the number of SC decoding attempts for different tree search methods proposed in this paper and compares it with that of the SCOS decoder in \cite{yuan_ML}. For $\texttt{TS-DFS-$\omega$}(\cdot)$, $\texttt{TS-BFS-$\omega$}(\cdot)$, $\texttt{TS-DFS-O-$\omega$}(\cdot)$, and $\texttt{TS-BFS-O-$\omega$}(\cdot)$ functions, $\omega=4$ is selected for $\mathcal{RM}(3,7)$, $\omega=3$ is selected for $\mathcal{RM}(4,7)$, and $\omega=5$ is selected for $\mathcal{RM}(2,8)$ to ensure near ML decoding performance. It can be seen that limiting the depth of the bit-flipping tree provides significant average complexity reduction for both DFS and BFS schemes. Moreover, the DFS scheme benefits from further average complexity reduction when the node visits are ordered. However, ordering the node visits results in negligible average complexity reduction for the BFS scheme. This is due to the fact that the BFS scheme naturally starts with smaller number of bit-flips. In fact, the BFS scheme achieves a lower average complexity than the SCOS decoder in \cite{yuan_ML}. Note that all the proposed decoders approach the complexity of a single SC decoding attempt for large values of $E_b/N_0$.

\section{Conclusion}

This paper proposed a tree search approach that achieves the ML decoding performance for RM codes. Two tree search methods are adopted: a DFS scheme that has a simple recursive structure; and a BFS scheme that results in a low decoding complexity. Several enhancements to reduce the worst-case and the average complexity of the proposed tree search approach are developed: 1) limiting the depth of the tree; 2) eliminating redundant node visits; and 3) ordering the node visits. Simulation results confirm the effect of the proposed methods to reduce complexity and show that the BFS scheme provides lower average decoding complexity than the existing tree search approach to decode RM codes.



\end{document}